\documentclass[conference]{IEEEtran}
\IEEEoverridecommandlockouts
\usepackage{cite}
\usepackage{amsmath,amssymb,amsfonts}
\usepackage{algorithmic}
\usepackage{graphicx}
\usepackage{textcomp}
\usepackage{xcolor}
\def\BibTeX{{\rm B\kern-.05em{\sc i\kern-.025em b}\kern-.08em
    T\kern-.1667em\lower.7ex\hbox{E}\kern-.125emX}}
\begin{document}

\title{Predicting direction of any stock index in a unified Transfer Learning Framework with an a Priori Causal Graph as main Input
\author{David Romain Djoumbissie\\
\IEEEauthorblockA{Diro, Université de Montreal; Canada Mortgage and Housing Corporation}\\
}
}


\maketitle
\begin{abstract}
We propose a unified multi-tasking framework to represent the complex and uncertain causal process of financial market dynamics, and then to predict the movement of any type of index with an application on the monthly direction of many important US market indices. We base our solution on three main pillars: (i) the combination of multidisciplinary knowledge (Financial economics, behavioral finance, market micro-structure and portfolio construction theories) to represent a global top-down dynamics of any financial market through a causal graph, (ii) the integration of forward looking unstructured data, different types of contexts (long, medium and short term) through latent variables/nodes and then, use a unique Variational autoencoder to learn simultaneously their distributional representation, (iii) and the use of transfer learning to share knowledge (Parameters, features representation and learning) between  all financial markets, increase the training sample, preserve the stability between training, validation, test sample and finally, use the fine-tuning to improve the predictive power on the specific market index. By example on the SP500 index, we obtain Accuracy, F1-score, and Matthew Correlation of 78\%, 76\% and 0.52 above the industry and other benchmarks on the test period of twelve years, which includes three unstable and difficult sub-periods to predict.
\end{abstract}


\section{Introduction}
Machine learning (ML) and more precisely deep learning (DL) has allowed, during the last decade, to obtain results that upset the state of the art and has become the dominant paradigm in various disciplines. The challenge of predicting the financial markets is not an exception \cite{TR10,WH18,Ca21}. In the asset management industry, many tasks use ML/DL, as the very short-term market prediction (minute/daily horizon) and the short/medium term prediction (monthly/quarterly horizon).\newline\\ 
According to many authors \cite{Pea18}, ML/DL is currently effective in learning trends, recognizing patterns and making predictions on stable data. However, it remains uncertain about interpretation, cause-and-effect relationships, and predictive power on unstable data.\newline\\
Many active portfolio managers have a short-term investment objective (one to 12 months) and need a dynamic decision-making framework with an ability, to explain, predict the future direction of the financial markets and above all, reconciling the prior domain knowledge with the process and output of different algorithms. The 
Tactical Asset Allocation (TAA) is a dynamic strategy that actively adjusts the view and the investment decision based on short-term market forecasts.\cite{ Ka21} identify more than one hundred well-established TAA funds with over 65 billion dollars in assets were attempting to deliver better risk-adjusted returns than static allocation funds. Each Manager use an investment decision process based on short-term prediction, and decide to buy/sell if prediction is up/down on a monthly/quarterly or annual basis. The cost of implementation (staff, computer, time, board acceptance, transaction cost) increases dramatically with the investment horizon and it is impossible to use the second / minute prediction to achieve the goal. On the other hand, many practices in the industry and studies \cite{Do11,Un21}  believes the prediction on the monthly horizon is well adapted for this goal. \newline\\
The dominant paradigm uses two main assumptions of market efficiency and rationality. The solution is transparent, easy to interpret, but the predictive power is poor. S. Hebert, holder of the Turing Prize (1975) and the Nobel Prize in economics (1978), had already emphasized the limits of pure rationality and the necessity to focus on limited and procedural rationality \cite{Ne59,Si82,Si91}.
More precisely, in a complex and uncertain environment, the investor is limited by a set of constraints, including: (i) time varying, imprecise and incomplete nature of the information; more around 80\% are unstructured \cite{Sq14,VU15}; (ii) a large number of cognitive biases \cite{KaTV74, KaTV79, KaTV98, Th05}; and (iii) the speed of analysis and interpretation of regularly updated information. The concepts of rationality and an optimal solution give way to the design and execution of a set of processes, integrating these limits.\newline\\ 
Over the last decade, many studies \cite{Ca21, Lu16,Po17,Pe17,BI17,DDR20,Ic17,Sg18,DI14,DI15,Chg17} have focused on the use of ML/DL, Bayesian and Knowledge-based Systems to provide alternative solutions. In order to improve the interpretation, explanatory and predictive power, different studies \cite{OBU19,WJ20} suggest: (i) the integration of forward-looking information and unstructured data; (ii) the consideration of hierarchical and indirect relationships; (iii) the use of prior knowledge and more complex algorithms to create features, learn representations, or directly represent and predict the market.\newline\\
The contrast between industry and academic studies, the lack of studies over long periods (covering the multiple shifts in market regimes), and the difference between the prediction on minutes/daily versus monthly/quarterly horizon, make the notion of the state of the art somewhat confused. Recent industry studies \cite{Ka21,BeJ17,KeA18} analyze the performance of various investment frameworks. The general conclusion is that active managers have failed to beat their benchmarks, especially over a long period. The benchmark is a passive framework with buy and hold the market index  (predict index is up all time). The assumption is, find a predicting process with better performance than a random walk \footnote{Stochastic or random process, that describes a path that consists of a succession of random steps.} is difficult.\newline\\
For monthly direction prediction, \cite{DDR20} identify five studies on S\&P500 and propose an informed machine learning solution that outperforms on the same metric and test period. We identify four main limitations of this solution:  i) The learning process is manual/ad hoc and the focus is on only one index;  ii) A very small number of risk drivers are used ; iii) The training sample is small (540 observations on 45 years), and the walk forward back-test approach doesn't solve the issue of unstable distribution (training vs test); and iv) finally, it still possible to improve the performance on the test period, especially during a volatile period.\newline\\   
For the authors in \cite{Ca93,We16}, transfer learning and a priori domain knowledge constitute a potential framework to go beyond those constraints. It allow simultaneously to learn many tasks, and propose a solution with the ability to identify and extract knowledge from many source task, and then apply it to another target task that shares commonalities. The top challenges to consider are:  identify a set of similar tasks; decide which knowledge to transfer (instances, features, parameters,...), and finally how to do the transfer.\newline\\ 
We propose a unified transfer learning solution, and use an a priori causal graph to represent the dynamic of any stock index . It allows us to address: i) The challenge to reduce the perception of "black box" around DL solutions, by propose a hybrid solution with an a priori graph in the top of the process, to encode a priori knowledge on causal interaction between any market index and the main drivers; ii) The issue of the poor predictive power on the task of monthly prediction with a small training sample; iii) The issues of noisy input and the difference between the distribution of the training versus the test sample; and iv) The issue of learning embedding on any market index simultaneously in a unified framework. To achieve this solution, our contribution focuses on introducing five main points that ensure the effectiveness of transfer learning on the tasks of representation and prediction on any financial markets:
\begin{itemize}
\item The identification of 235 source tasks (monthly prediction of 235 market indices) alongside the target task of predicting any specific market index.
\item The transformation of raw input in a new space (from noisy input to stable features). At the end of each month, we represent each index by a set of 20 latent context variables with 125 attributes for each, and allow to share feature between task by clustering approach.
\item The proposition of a hierarchical causal graph with 20 nodes/latent context drivers. It allows to represent the dynamic causal process of any market index, encode all interactions identified in financial/economic theories between all nodes, and between each node and long/medium/short term risk factors.  
\item A unified framework to merge all sample and simultaneously learn nodes representation (transfer parameter/feature) via a unique Variational autoencoder (VAE).
\item A validation framework over a long period, with sub-periods known to be very difficult to predict. We use the monthly prediction of few indices as a target task in a two-step process: pre-train a classifier on all 235 source tasks, with node embedding as input, and then train the final classifier (transfer parameter and fine-tuning) only on the target sample.
\end{itemize}
The rest of the article is organized into 4 sections. In section 2, we review related work on the stock index prediction. Section 3 describes our solution through an a priori causal graph , the feature engineering and the learning process. We present the experiments and empirical results in section 4. Finally, the last section provides the conclusion and future work.
\section{Related Works}
During the last 50 years, the main paradigm has remained based on a rational framework, with a desire to understand and explain all the process from input transformation, feature engineering, cause-effect analysis to interpretation of output. Various studies \cite{SH64,FA91,FA93,Ca14} use a priori knowledge to select a few variables for an econometric model. The solution makes it possible to explain ex post the different interactions with the main drivers. However, the predictive power is poor, especially during volatile market regimes, and it is difficult to beat the random walk on the task of predicting the monthly/quarterly direction of the market index.\newline\\ 
The focus today is to use the DL \cite{Ca21} to learn representation and predict the stocks indexes. \cite{OBU19} identify 140 studies (2005-19) with only 9 on monthly/quarterly horizon and the test period less than 5 years. DL are not able to surpass results obtained by traditional ML on monthly horizon, but were generally better than traditional ML on the daily/intraday horizon. \cite{WJ20} identify 124 studies (2017-19), with 106 on a daily horizon and 19 on an intraday horizon, where DL improves the predictive power.\\ The predictive power on the very short horizon is encouraging. However, more than 95\% of the studies propose a minute/daily prediction because the size of training sample is high (collect data every minute vs month). Most solutions involve tests over a very short period (less than 4 years), so it is difficult to generalize the conclusion in a unstable environment with frequent shifts in the market regime. In addition, predicting over a very short horizon solves a specific type of problem in the industry (trade optimization,...). Nevertheless, it is different from the issue of monthly prediction for tactical decisions with an investment horizon around one to 12 months. Lastly, most of the solutions put the focus on predictive power, but the main battle for acceptance in the industry are the lack of interpretation (representation/prediction), and the capacity to reconcile domain knowledge and cause-effect analysis.\newline\\
For monthly prediction, \cite{DDR20} identifies five studies on predicting the specific and most important index on US market (S\&P500) and proposes a benchmark for solution with a focus on interpretation, explanatory/predictive power. This solution uses informed machine learning through a three-step manual/ad hoc process and we use as a benchmark.
\section{The Proposed Framework}
We tackle the monthly prediction of the direction of S\&P500 Index and any other market index, as a binary classification task in a homogeneous multitask and transfer learning (MTL/TL) framework depicted in Figure 1. The solution puts the focus on a set of homogeneous tasks and have four main specificities: the creation of a set of adapted features in a new space, the combination of the training samples of all the target/source tasks, the deployment of a simultaneous learning process on a large/stable sample of features, and finally the improvement of learning process on the target task. The challenge consists in identifying a set of homogeneous tasks and address three main issues: what knowledge to share ? When and how to share ?\newline\\
The S\&P500 is the most important index on the US market, and our ultimate goal is to improve its monthly prediction (Up/Down) and go beyond with any other index in the same framework. We use the domain knowledge to identify 235 related indices and cover all sectors/sub-sectors, asset classes and risk factors on the US markets, then formulate and solve the monthly prediction task in a homogeneous MTL/TL on 235 indices. We organize our solution around two steps :\newline
\begin{figure}
\includegraphics[width=0.8\textwidth,height=0.45\textheight]{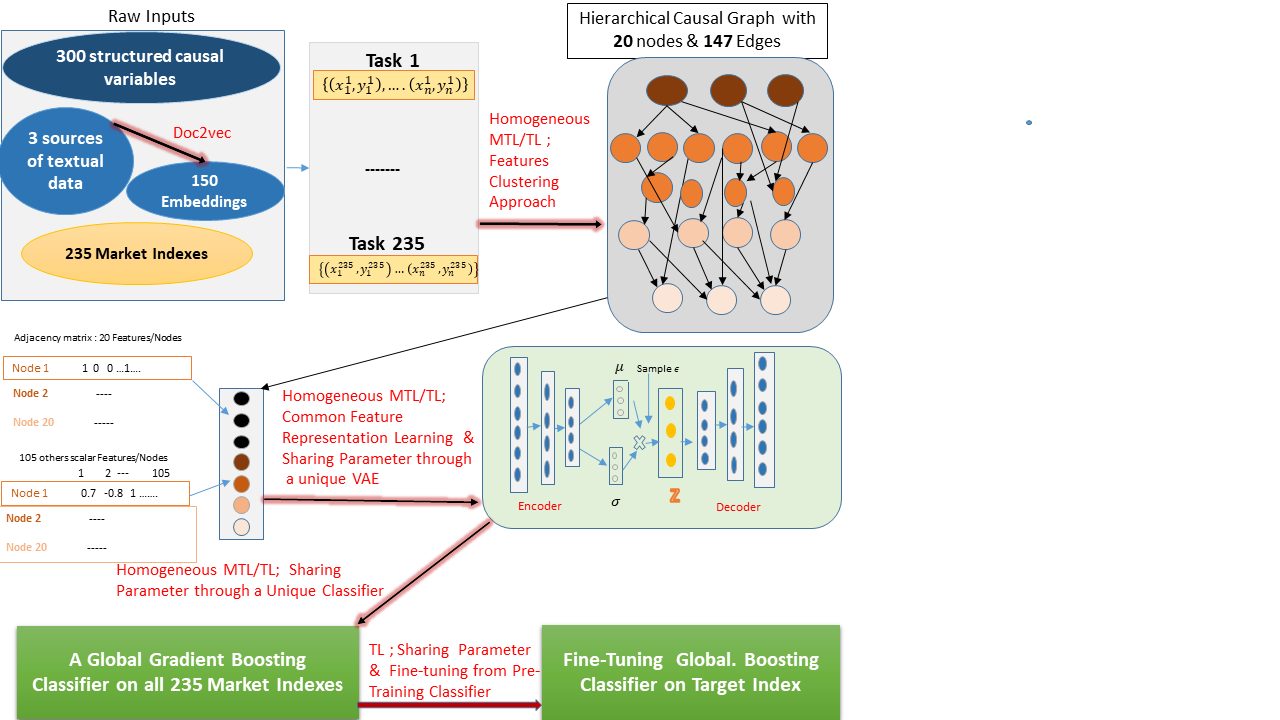}
\caption {\scriptsize Transfer 
Learning framework for market index representation/prediction with a focus on upstream Interpretation}
\end{figure}
\begin{figure}
\includegraphics[width=0.55\textwidth,height=0.45\textheight]{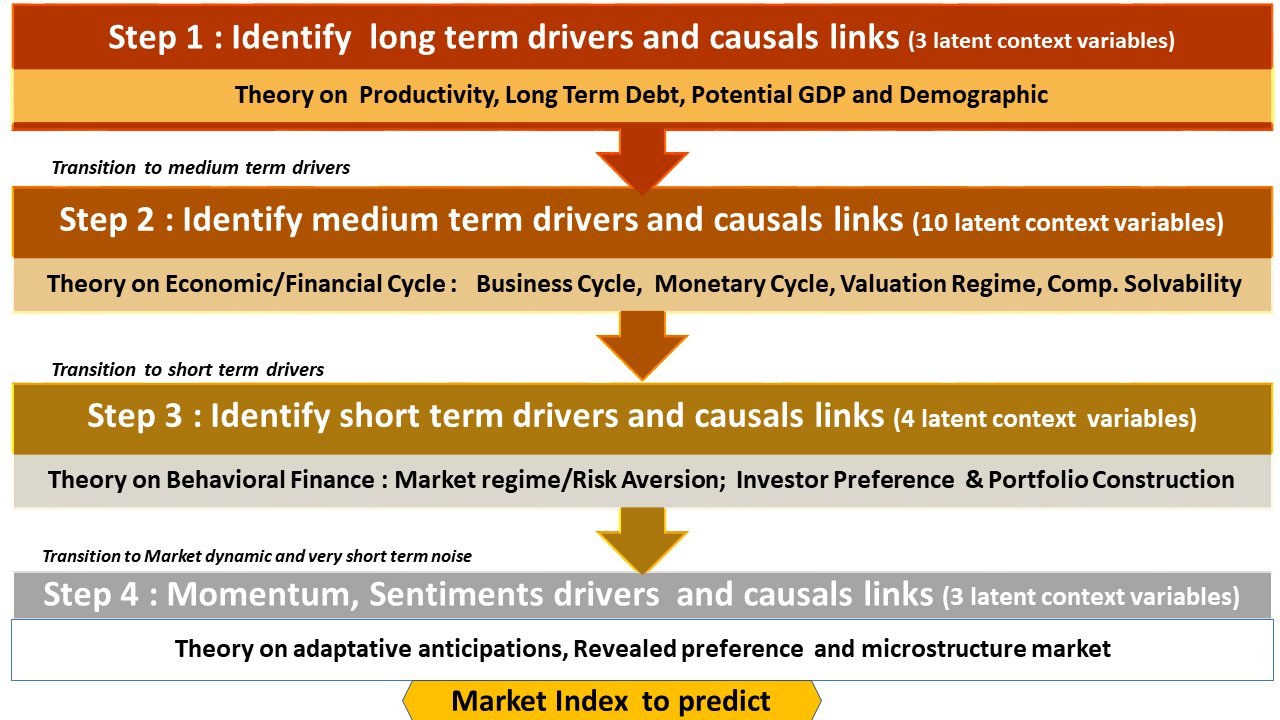}
\caption {\scriptsize Main Component of Causal hierarchical top down dynamic of any Market Index }
\end{figure}
The first step (Section A), allows the transformation of the raw input into a new stable feature space (share feature by clustering approach), through an a priori causal graph with 20 nodes or well-known latent context risk factors, 147 oriented edges and 125 features/attributes for each node. Using a feature clustering approach, those 125 features allow a summary of the dynamic and hierarchical interaction between any market index and a set of 20 latent factors.\\
The second 3-step learning process (Section B and C) allows the learning of feature representation, via the same VAE for all the 235 tasks (sharing feature and parameter by encoding approach). Then, we pre-train a global classifier on the merged sample of Embeddings on all tasks (sharing parameter by gradient-boosting and learn only one classifier). The Final step (transfer parameter via model selection approach and fine-tuning) uses the hyper-parameters/parameters of a pre-trained decision tree, to initialize and train a final classifier only on the target task (Ex: S\&P500).\newline
We formalize the learning part with four equations below :\\
Assuming three indexes,\\
\textbf{t} $ \in \{1,...576\}\;$: Monthly time index between 1970-2018; \\ \textbf{i} $ \in \{1,...235\}\;$: Market indices for source/target task; \\
\textbf{k} $\in \{1,...20\}\;$: 20 nodes of a graph \textbf{$G_{i,t}$}; each node is represented by 125 attributes \textbf{$F_{t}^{i,k}$}, and allows the encoding of a set of dynamic information that represent, an interaction between a market index i and node k at time t, then the existence of a causal link from node k to the rest of nodes.
\begin{eqnarray}
 F_t^{i,k}= h(Raw\_Input_\textbf{t}^{i,k}; Raw\_Input_\textbf{{1:t}}^{i,k}) \\
 E_t^{i,k}= g(F_t^{i,k}) \\ 
 y^i_{t+1}= f_{Global\_Classifier}(E_t^{i,1},...E_t^{i,20}) \\
y^{Target}_{t+1}= f_{Target}(E_t^{Target,1},..,E_t^{Target,20})
\end{eqnarray}
$Raw\_Input_\textbf{t}^{i,k}$ or $Raw\_Input_\textbf{{1:t}}^{i,k}$ : Sub-set of raw input used to measure a key drivers k at t, or period 1 to t and adjusted/organized based on information from market index i.\newline\\
\textbf{h}: Similarity function used to measure proximity  between a vector at t versus a set of realization for period 1:t or any sub-set of realization organize in a specific cluster.\newline \\ 
\textbf{F}$^{i,k}_{t}$ : At time t, a set of 125 features/attributes of node k in the graph \textbf{$G_{i,t}$}. It allow to summarize all interaction between a latent context or key driver k, a market index i and other nodes.\newline\\
\textbf{E}$^{i,k}_{t}$: At t, a latent vector (d dimension) from embedding of node k with a set of attribute \textbf{F}$^{i,k}_{t}$.\newline \\
\textbf{g}: A Variational autoencoder. The input is  \textbf{F}$^{i,k}_{t}$ and the output is \textbf{E}$^{i,k}_{t}$ (from encoder). We train the VAE to minimise the error between \textbf{F}$^{i,k}_{t}$ and its estimate  $\widehat{\textbf{F}^{i,k}_{t}}$.\newline \\ 
\textbf{f}$_{Global\_Classifier}$: A classifier (gradient boosting tree), the input is the vector of d*20 scalars $(E_t^{i,1}..E_t^{i,20})$ and the output is 0/1 (market index i is up/down and i=1...235).\newline \\
\textbf{y}$^i_{t+1}$: 0/1 if the monthly price direction of the market index i is Up/Down during the month t to t+1.\newline\\ 
\textbf{f}$_{Target}$: A Classifier (gradient boosting tree), fine-tuning from the  \textbf{f}$_{Global\_Classifier}$ on the target sample. The output is \textbf{y}$^{Target}_{t+1}$ (0/1) and the input is  $(\textbf{E}_t^{Target,1},...,\textbf{E}_t^{Target,20})$. The target could be any index to predict. 
\subsection{\textbf{Description of the a priori causal graph, the raw inputs and the feature engineering process}}
\subsubsection{\textbf{The a priori causal graph}} More formaly, 
let $G_{it}=< N,\xi,({F}^{i,1}_{t},..{F}^{i,20}_{t})>$. At the end of month t, We describe the causal process of the dynamics of any market index i through a causal graph. The graph has three main components: i) a set of raw inputs, 300 observable causal backward looking drivers, as well as 150 ones from forward looking unstructured data; ii) 20 nodes (N) where each has  
125 attributes (${F}^{i,k}_{t})$ and summarize all information on the interactions (direct/hierarchic, linear/non-linear, static/time-varying) between a market index i and the short/medium/long term drivers (e.g: economic cycle, risk aversion or valuation regime); iii) 147 edges ($\xi$) that reflect, the existence of a direct causal link between two latent context drivers or nodes.\newline\\
The a priori graph structure lies on two main source of knowledge. Our 17 years of experience on the market and over 50 years of literature \cite{KaTV74, KaTV79,KaTV98,Th05,FA89,RA21,LO21,BA96,LU18,GI02,RO11,RO16} on the financial markets (financial economic theory, behavioral finance, fundamental analysis, micro-structure and technical analysis). We describe a global hierarchical top down interaction between long, medium, short term drivers and any market index (Figure 2). Then, the 20 nodes with associated raw inputs are described in Table~\ref{Table 1}. We suggest \cite{DE17} for more details on graphical model and potential applications in finance.\\
\subsubsection{\textbf{The Raw Inputs}} 
We have 3 categories of raw inputs:\\
\textbf{Market Indices:}\\
Let $\textbf{I}^i_{1:t}$: vector of 12096 daily level of index i (1970-18). We select 235 indices for source/target tasks based on three main criteria: 
\begin{itemize}
\item  The statistic properties, the regularity of the movement in each of the market regime and the business cycle. \item Investors use these indices to implement specific strategies and investment objectives.
\item Availability of long history (K. French website) \footnote{\scriptsize  mba.tuck.dartmouth.edu/pages/faculty/ken.french}.
\end{itemize}
We select 2 main US indices, 49 industries and 11 sectors (Global Industry Classification Standard), 173 indexes (Risk Premium,...).\newline \\
\textbf{Risk Drivers from structured data (Table1) :}\\
We select 300 variables identified in the literature as a long/medium/short term drivers. Ex: company earning, volatility curve , productivity.\\
$\textbf{V}_{1:t}$: Matrix of 12096$\times$300 scalars, daily realization 1970-18 .\newline\\
\textbf{Risk Drivers from Unstructured data (Table1) :} \\
We select 3 sources of forward looking unstructured drivers. The Beige Book, the report on current monetary policy from the Central Bank; a set of article on the main fact on economic recession. A doc2vec algorithm allow to transform each document to a set of 25/50 (hyper-parameters) embeddings. We combine all of them and : \\
$\textbf{V}^{i,k}_{1:t}$: Matrix of 12096$\times n_{k}$ scalars or daily observation of raw inputs between 1970-18. It is a set of $n_{k}$ inputs (sub-set of 450 raw inputs) identified in the literature as a main variable to characterize the latent context/main key driver k (table1).\\
\subsubsection{\textbf{Features engineering by clustering approach (mahalanobis/Cosinus Similarity)}} 
More than 50 years of literature on the modelization of the financial market \cite{RA21,LO21,RO16,VO10}  allow now to identify a relative consensus around three main empirical facts. i) Market prediction is a big challenge, it is basically a complex process with many sources of uncertainties. ii) the dynamic of a financial market could be summarize through a sequence of three main regimes. The bull, the range bound and the bear market. Ex-ante, it is difficult to anticipate but ex-post, a set of rulers or algorithm can allow to identify the end and the beginning of each regime. ii) the dynamic of economic could be summary through of a different cycle, from 3-5 years to very long term cycle of 40 or more years (Kondratiev waves).
\newline \\
The capacity to identify the right current market regime and current economic cycle are fundamental for the success of any tactical/dynamic investment process. It will be the main motivation of all our Features engineering step.\\
Let $\textbf{F}_{t}^{i,k}$ =  $\{{C}_{t,i,k}^{1},{C}_{t,i,k}^{2},{C}_{t,i,k}^{3},{C}_{t,i,k}^{4},R^i_{(t-21:t)}, A^k\}$, a set of 125 features. For each index i, we use empirical studies on market regime to organize historical raw inputs into (m= 4) clusters \textbf{(1: Bullish market regime, 2: Range Bound, 3: Bearish, 4: current long-term economic cycle)} .\newline\\
Let $\tiny{I^i_{t_0},...,I^i_{t_n}}$ be, the sequence of level of index i, observed between $\tiny{t_0},...{t_n}$. Ex-post, the market index i was in a \textbf{bullish regime} during $\tiny{t_0}...{t_h}$, if starting at ${t_0}$, the index i rises gradually above a certain threshold without returning below the initial price at ${t_0}$. Meaning the set of points: $\{t_0..,t_j,..t_h;\;0\leq j\leq
h\; and\; \tiny{I}^i_{t_j}\geq \tiny{I}^i_{t_0}\,\;\&\;\,\tiny{I}^i_{t_h}\geq (1+\lambda)\tiny{I}^i_{t_0} \}$; $\;\lambda$: Hyper-parameter.\newline 
The \textbf{bearish regime} is exactly the opposite of  the bullish regime. The \textbf{Range Bound regime} is obtained when the price oscillate between up/down, but stay under/above the threshold when up/down.\newline
These three regimes are based on what financial industry call short term market regime, while the last one, \textbf{Current long term cycle} is a assumption on that, the period from 1970-2018 was in the same long term economic cycle.\newline\\  
$\mu_{tik}^{m}$: Vector of dimension $n_{k}$ (size raw input for latent variable k). At t, a moving average of each raw input $V^{ik}_{(1:t)}$ when the Market Index i is in one of the market regime m;\newline\\
$\textbf{S}_{tik}^{m}$: Matrix of dimension $n_{k}\times n_{k}$  designating the moving covariance matrix of raw input $V^{ik}_{(1:t)}$ (idem as $\mu$).\newline\\   
$\textbf{C}_{tik}^{m}$\;=\;\textbf{h}{$(V^{ik}_{(t-21:t)} ,\mu_{tik}^{m},\frac{1}{S_{tik}^{m}})$}: Vector of 21 features obtained by similarity between the last month (21 market days)
$V^{ik}_{(t-21:t)}$ and its past realizations in the cluster m ($\mu_{tik}^{m}\;,\;S_{tik}^{m}$).\newline
\textbf{h} : Mahalanobis similarity (structured data) and Cosinus similarity (Unstrured data). The mahalanobis is very easy to interpret in the context of financial market dynamic (co-mouvement and covariance between asset), and allow to identify potential regime shift in the multivariate data. \\
\textbf{h}: $R^{n\times21}\times R^n \times R^{n\times n} {\longrightarrow}\;R^+$ \\
{\tiny$(V^{ik}_{(t-21:t)},\mu_{tik}^{m},\frac{1}{S_{tik}^{m}})\tiny\longrightarrow\sqrt{(V^{ik}_{(t-21:t)}-\mu_{tik}^{m})^t\times\frac{1}{S_{tik}^{m}}\times(V^{ik}_{(t-21:t)}-\mu_{tik}^{m})} $}\\
$\textbf{R}^i_{(t-21:t)}$ : At t, the daily variation of market index i ($I^i_{(t-21:t)}$) during the last month or 21 market days.\newline
$\textbf{A}^{k}$ : Vector of 20 Boolean from adjacent matrix for latent variable k (edge or causal interaction between nodes).
\subsection{\textbf{Nodes embedding through VAE\newline}}
In the first step, a predefined cluster and similarity measure (share feature by Clustering approach) allow to create a new space with more stable features. Then, we learn the representation of features with the same VAE and use its as input for the classifier.\newline\\
At the end of each month t, the input of the VAE is a node with 125 features or attributes and the output (d embedding) characterize the current dynamic of each node is derived by minimizing a loss function L with 3 components. The square error, the Kullback divergence for 105 continuous feature (C), and the binary cross-entropy for the 20 discrete feature (D).\newline
 L =  $\|F_{t,C}^{i,k} - \widehat{F_{t,C}^{i,k}} \|^2$ – $D_{KL}(N(E_t^{i,k}|F_{t,C}^{i,k})|| N(0,1))$\\ + $ \sum_{k=1}^K((F_{t,D}^{i,k}*log(\widehat{{F_{t,D}^{i,k}}}))+(1-F_{t,D}^{i,k})*log(1-\widehat{{F_{t,D}^{i,k}}}))$\newline\\
We try different structure, Convolutional (CNN), Recurrent (GRU) and Feedforward (FF), size of Network (3, 6 or more hidden layers) and embedding dimension (5, 10). 
\subsection{\textbf{Prediction target index via transfer parameter from a pre-trained classifier on 235 indices\newline}}
Our homogeneous MTL/TL framework allows in the precedent step, to merge all embedding for 20 nodes on the overall training sample (235 tasks of market index prediction). Then, we use gradient boosting algorithm and share parameter between tasks by pre-training only one Global Classifier. After that, we use a cross validation (share parameter based on model selection approach and fine-tuning) to select the first set of hyper-parameter/parameter and then, finalize the training only on the target sample.\newline\\ 
The input for a Global Classifier  ({f}$_{Global\_Classifier}$) is the set of d*20 embedding on the sample of 235 market indices $(\textbf{E}_t^{i,1},...,\textbf{E}_t^{i,20})$ and the output is $y^i_{t+1}$.\newline\\ 
For the final classifier on target task ($f_{Target}$), We initialize with the pre-trained Global Classifier, the input is the set of d*20 embedding $(\textbf{E}_t^{target,1},...,\textbf{E}_t^{target,20})$ but only on the target sample, and the output is $y^{target}_{t+1}$. We apply on 6 of the most important market/sector indices on the US.The loss function is : 
L =$ -\sum_{t=1}^T(y_{t}^{target}*log(odd))-log(1-p)$.\\
(odd = p/(1-p) and p: predicted probability).
\section{\bf Experiments and Empirical Results}
\subsection{\bf Training and validation protocol}
We train the model on a sample covering different market regimes. Then test it, and make sure the model improves predictions during both stable and unstable market regimes.\newline\\
The VAE is trained on the sample of 1128000 nodes (240 months $\times$ 235 indexes $\times$ 20 nodes) from 1972 to 1992, and the Encoder allows to generate embedding over the period 1993 to 2018. We train a global classifier on all sample of embedding over the period 1993 to 2004 (240 months $\times$ 235 indexes) and use the early stopping approach to select optimal parameters on the validation sample (2005-2006). After that, we update the hyper-parameters/parameters only on the target sample (Ex: S\&P500). \newline\\
We use the last 12 years (2007-2018) for out of time sample test on the different target task.
\subsection{\bf API and Hyper-parameters selection} 
We use Python 3.6, various APIs such as CNN/GRU, Doc2vec algorithms on Tensorflow, keras and Gensim, then GridSearchCV and LGBMCClassifier. In the first step, we use domain knowledge to provide the number of market regimes. In the second step, we choose the set of hyper\-parameters to maximize the discriminating power over the training, validation period. These include: the architecture of VAE, the dimension of node Embedding, and the classifier hyper-parameters/parameters (depth of the trees, learning rate,…). 
\begin{table}[t!]
\caption {\label{Table 1} Latent Context Drivers and Raw Inputs} 
\begin{tabular}{|p{3.5cm}|p{5.5cm}|}
\hline
\multicolumn{2}{|c|}{\scriptsize \bf 20 Latent Context Variables} \\
\hline
{\scriptsize \bf Long-Term Latent D.}&{\scriptsize \bf Raw Inputs}\\
\hline
\tiny - Productivity &\tiny Total Labor Prod.,Real Output Per Hour (Business, Non Financial, Non Farm).  \\
\tiny - Long Term Debt &\tiny Long-Term Debt Securities (Gvnt, Non Profit Organisation, Private Institution, Inv. Fund, Pension Fund, Monetary Autority, Insurance Comp., Local Gvnt, Other).  \\
\tiny - Short Term Debt &\tiny Short-Term Debt Securities (.....). \\
\hline
{\scriptsize \bf Cyclical Latent D.}&{}\\
\hline
\tiny - Monetary Cycle &\tiny M1,M2,Fed Funds,MZM Money Stock  Bank Prime Loan Rate, 3 month T. Bill, 6 month T. Bill, 12 month T. Bill, Finance rate on personal loan 24 months, Finance rate on consumer loan. \\
\tiny - Rate Term Structure Regime &\tiny Fed Fund, 3, 6, 12 month T. Bill, 1, 3, 5, 7, 10-Years T. constant maturity rate, Moody's Seasoned Aaa Corporate Bond Yield, Moody's Seasoned Baa Corporate Bond Yield, 30-Year Fixed Rate Mortgage Average in the United States.\\ 
\tiny - Eco Business Cycle&\tiny Industrial Production, Inflation\\
\tiny - Market Business Cycle&\tiny Return of well know 7 Risk Premium ( Equity premium, Cyclical, Value, Term, Credit, Carry, Safe Haven)\\
\tiny - Household Solvency Regime &\tiny Net worth level, Disposal personal income, Owners' equity in Real Estate, Consumer credit, Res. morgage, total liability, commercial morgage liability, other loans)\\
\tiny - Valuation Regime &\tiny Price Earning (curve many horizon)\\
\tiny - Monetary Policy Regime from unstructured data&\tiny Corpus of FOMC \\
\tiny - Business Cycle from Leading unstructured Data&\tiny Corpus of Beige Book \\
\tiny - Recession from FED&\tiny Corpus of Recession Knowledge from Academic and Central Bank Article and Corpus of FOMC \\
\tiny - Recession from Business Community&\tiny Corpus of Recession Knowledge from Academic and Central Bank Article and Corpus of Beige Book \\
\hline
{\scriptsize \bf Short-Term Latent D.}&{}\\
\hline
\tiny - Co-movement Regime on cross-asset Return of 70 market indices  &\tiny - Fixed Income, Commodities, Currency, Equity and 49 Industrial indices from GICS.\\
\tiny - Co-Volatility Regime cross-asset volatility of 70 market indices&\tiny ---.\\
\tiny - Co-Skew Regime on cross-asset Skewness of 70 market indices&\tiny ---.\\
\tiny - Co-Kurtosis Regime cross-asset Kurtosis of 70 market indices&\tiny ---\\
\hline
{\scriptsize \bf Very Short-Term D.}&{}\\
\hline
\tiny - Momentum Return regime of market index to predict&\tiny curve of return different horizon\\
\tiny - Momentum Volatility regime of market index to predict&\tiny volatility curve different horizon\\
\tiny - Risk Aversion Regime &\tiny Risk Premium Curve of Main well know Risk Factor or return Risk factor minus Risk Free rate (equity, Sector, Fama French factor,...)\\
\hline
\end{tabular}
\end{table}
\subsection{\bf Evaluation metrics and main analysis} 
We use three adapted metrics\cite{Ic17}. The Accuracy \footnote{the total percentage of good predictions (up/down)} (\textit{ACC}), the F1-score\footnote{the harmonic average between Precision and Recall} and the Matthews Correlation Coefficient\footnote{Correlation between the observed and predicted binary classification} \textit({MCC}). F1 and MCC allow a relevant analysis of the cost of errors. Indeed, the cost of bad decisions is high and the biggest challenge is to have models with good \textit{ACC}, but especially an ability to limit false positives and false negatives (predict down when market is up or predict up and the market is down).\\
We experiment with a focus on a few main points:
\begin{itemize}
\item Consider the main US market index (S\&P500) as a target, and be able to compare with two benchmarks (the industry and the best model on monthly prediction of S\&P500 proposed in \cite{DDR20}), over the test period (2007 to 2018) and three unstable sub-periods (2007-08, 2011-12, 2015-16), which is difficult to predict. We also evaluate the impact of : i) the VAE network architecture (CNN, GRU network) and the size (3, 6 or 9 hidden layers), ii) then the dimension of each latent context variable or nodes embedding (5 versus 10). \item Consider the best solution and apply on 6 of the most popular Market/Sector indices as target index to predict (S\&P500, Financial, Technology, Industrial, Material, Consumer stable sector) .
\end{itemize}
\subsection{\bf Performance and comparison\newline}
The empirical output on Table~\ref{Table 2} and Table~\ref{Table 3} provides strong support for the validity of our framework. We improve three key metrics (ACC, F1-Score and MCC) over a test period (2007-2018), and obtain stable output on 3 sub-period considered as volatile and quite difficult to predict. On the target task of predicting the main Market index S\&P500, our solution outperforms the industry benchmark and the best model in the study \cite{DDR20} on the same validation/test period. Precisely:
\begin{itemize}
\item We try different type of VAE network (simple FF, GRU and the combination of CNN/GRU ). The best results are obtained with d equal 10 and 5, the CNN/GRU network with 3 and 6 layers  and the  set of hyper-parameters (LR: 0.00475, max\_deph: -1, max\_bin and max\_leaves: 512).
\item  The Table~\ref{Table 2} and Table~\ref{Table 3} confirms the advantage of combining CNN with GRU and shows that:\\  
i) Over the entire test period (2007-2018), the CNN/GRU VAE network with 3 layers and embedding of dimension 5 outperforms; ii) Combining CNN with a GRU architecture generally improves performance over simple GRU or a FF network. iii) Increasing the number of hidden layers do not necessary improve the performance.
\item For the Global Classifier before fine-tuning on all the 235 tasks (Table~\ref{Table 2}), the CNN/GRU VAE with 3 hidden layers and 5 embedding per node outperforms with a stable and high performance on the validation set (ACC, F1 and MCC are respectively 78\%, 76\%, 0.52), and the test set (ACC, F1 and MCC are respectively 76\%, 76\%, 0.51). 
\item We initialize the new classifier with hyper-parameters from Global Classifier. Then, a fine-tuning (training on target sample, more small learning rate and max\_leaves) allow to improve the performance on five of the most important sector index as the target task. We improve all metrics (ACC, F1, MCC) on the S\&p500 market index and 4 sectors indices (Technology, Industrial, Material, Consumer Stable). The three metrics before the fine-tuning and after the fine-tuning for the 6 indices are available on Table~\ref{Table 2}.
\item Over the entire test period (2007-18) on the target task of S\&P500 (Table~\ref{Table 3}), our model outperforms the two benchmarks on the three metrics. The ACC, F1 score and MCC are respectively 78\%, 76\%, 0.52 compare to 70.1\%, 61\% and 0.30 for the best model in \cite{DDR20}. The benchmark of the industry are respectively 63\%, 38\% and 0 respectively.  
\item On the three volatile sub-periods (Table~\ref{Table 3}), our solution keep the performance stable and relatively high on all metrics (above 71\% on ACC and 63\% on F1 score). The performance of the two benchmarks decrease with the ACC under 59\% on two sub-periods. The industry benchmark (buy index and hold or suppose index is up all time) confirms that the approach can be sustainable during stable periods, but create enough damage as seen in the markets during volatile periods.
This confirm what we know in the industry, one of the big challenge is not only in the capacity to have high ACC, but to keep high F1 score and ultimately be able to preserve this high score during the volatile market. 
\end{itemize}
We can conclude that, our framework improves the stability, outperforms over all the test periods, and does better over all the three unstable sub-periods. 
\begin{table}[t!]
\caption {\label{Table 2} \scriptsize \bf Monthly prediction on Train(1993-04), Validation (2005-06) Test period (2007-18)} 
\begin{tabular}{|p{3.5cm}|p{0.25cm}|p{0.25cm}|p{0.25cm}|p{0.25cm}|p{0.25cm}|p{0.25cm}|p{0.25cm}|p{0.25cm}|p{0.25cm}|}
\hline
\multicolumn{10}{|c|}{\tiny  \bf Global Classifier on All 235 Index Before fine-tuning on target sample}\\
\hline
Model &
\multicolumn{3}{c|}{Train} &
\multicolumn{3}{c|}{Validation} &
\multicolumn{3}{c|}{Test} \\
\hline
&{\tiny \bf ACC}& {\tiny \bf F1} &{\tiny \bf MCC}&{\tiny \bf ACC}& {\tiny \bf F1} &{\tiny \bf MCC}&{\tiny \bf ACC}& {\tiny \bf F1} &{\tiny \bf MCC}\\
\hline
\tiny Benchmark Industry(Index up all time) &\tiny &\tiny&\tiny&\tiny61&\tiny37&\tiny0&\tiny43&\tiny30&\tiny0\\
\tiny VAE 3 Layers (GRU), 5 Embeddings/per Node  &\tiny96&\tiny96&\tiny0.91&\tiny66&\tiny62&\tiny0.25&\tiny72&\tiny72&\tiny0.44 \\
\tiny VAE 6 Layers (CNNGRU), 10 Embeddings  &\tiny93&\tiny93&\tiny0.85&\tiny69&\tiny65&\tiny0.32&\tiny64&\tiny64&\tiny0.30 \\
\tiny VAE 3 Layers (CNNGRU), 10 Embeddings &\tiny99&\tiny99&\tiny0.97&\tiny74&\tiny66&\tiny0.45&\tiny71&\tiny71&\tiny0.43 \\
\tiny VAE 3 Layers (CNNGRU), 5 Embeddings  &\tiny \bf 99&\tiny \bf 99&\tiny \bf 0.99&\tiny \bf 78&\tiny \bf76&\tiny \bf 0.52&\tiny  \bf 76&\tiny \bf 76&\tiny \bf 0.51 \\
\hline
\multicolumn{10}{|c|}{\tiny  \bf Best Model ( VAE 3 layers, 5 Emb.) on 6 Main US market index as Target : Before and after fine-tuning on target sample}\\
\hline
&{\tiny \bf ACC}& {\tiny \bf F1} &{\tiny \bf MCC}&{\tiny \bf ACC}& {\tiny \bf F1} &{\tiny \bf MCC}&{\tiny \bf ACC}& {\tiny \bf F1} &{\tiny \bf MCC}\\
\hline
\tiny Benchmark Industry(SP500 up all time) &\tiny &\tiny&\tiny&\tiny66&\tiny40&\tiny0&\tiny63&\tiny38&\tiny0\\
\tiny Best Model before Fine-Tuning. SP500 Index &\tiny 99&\tiny99&\tiny0.97&\tiny71&\tiny63&\tiny0.29&\tiny77&\tiny75&\tiny0.49 \\
\tiny Best Model after Fine-Tuning (F-T). SP500 Index &\tiny 99&\tiny99&\tiny0.98&\tiny79&\tiny74&\tiny0.50&\tiny \bf 78&\tiny \bf 76&\tiny \bf0.52\\
\hline
\tiny Benchmark Industry(Technology up all time) &\tiny &\tiny&\tiny&\tiny46&\tiny31&\tiny0&\tiny62&\tiny38&\tiny0\\
\tiny Best Model before F-T, Technology Sector Index  &\tiny 96&\tiny96&\tiny0.91&\tiny67&\tiny64&\tiny0.47&\tiny77&\tiny75&\tiny0.50 \\
\tiny Best Model after F-T, Technology  Sector Index &\tiny 96&\tiny96&\tiny0.91&\tiny79&\tiny79&\tiny0.65&\tiny \bf 79&\tiny \bf 77&\tiny \bf0.55\\
\hline
\tiny Benchmark Industry(Industrial Index up all time) &\tiny &\tiny&\tiny&\tiny67&\tiny40&\tiny0&\tiny65&\tiny39&\tiny0\\
\tiny Best Model before F-T, Industrial Sector Index  &\tiny 97&\tiny97&\tiny0.94&\tiny75&\tiny67&\tiny0.39&\tiny74&\tiny72&\tiny0.43\\
\tiny Best Model after F-T,  Industrial Sector Index  &\tiny 99&\tiny98&\tiny0.97&\tiny79&\tiny76&\tiny0.51&\tiny \bf 75&\tiny \bf 73&\tiny \bf0.46\\
\hline
\tiny Benchmark Industry(Financial Index up all time) &\tiny &\tiny&\tiny&\tiny70&\tiny41&\tiny0&\tiny56&\tiny36&\tiny0\\
\tiny Best Model before F-T, Financial Sector Index  &\tiny 98&\tiny98&\tiny0.95&\tiny71&\tiny59&\tiny0.20&\tiny78&\tiny77&\tiny0.54 \\
\tiny Best Model after F-T, Financial Sector Index &\tiny 98&\tiny98&\tiny0.95&\tiny75&\tiny67&\tiny0.34&\tiny 78&\tiny 77&\tiny 0.54\\
\hline
\tiny Benchmark Industry(Material Index up all time) &\tiny &\tiny&\tiny&\tiny50&\tiny33&\tiny0&\tiny54&\tiny35&\tiny0\\
\tiny Best Model before F-T, Material Sector Index  &\tiny 97&\tiny96&\tiny0.93&\tiny75&\tiny74&\tiny0.53&\tiny72&\tiny71&\tiny0.42\\
\tiny Best Model after F-T, Material  Sector Index &\tiny 99&\tiny99&\tiny0.98&\tiny75&\tiny74&\tiny0.53&\tiny \bf 76&\tiny \bf 75&\tiny \bf0.50\\
\hline
\tiny Benchmark Industry(Cons. Stable up all time) &\tiny &\tiny&\tiny&\tiny71&\tiny41&\tiny0&\tiny60&\tiny37&\tiny0\\
\tiny Best Model before F-T,Consumer Stable Index  &\tiny 94&\tiny93&\tiny0.86&\tiny75&\tiny62&\tiny0.31&\tiny71&\tiny69&\tiny0.38 \\
\tiny Best Model after F-T, Consumer Stable Index &\tiny 95&\tiny95&\tiny0.89&\tiny88&\tiny84&\tiny0.68&\tiny \bf 72&\tiny \bf 70&\tiny \bf0.41\\
\hline
\end{tabular}
\end{table}
\begin{table}[t!]
\caption {\label{Table 3} \tiny \bf Monthly prediction with Market Index S\&P500 as Target and comparison with 2 Benchmarks} 
\begin{tabular}{|p{3.75cm}|p{0.75cm}|p{0.751cm}|p{0.75cm}|}
\hline
\multicolumn{4}{|c|}{\tiny \bf All test Period 2007-2018} \\
\hline
&{\tiny \bf ACC}& {\tiny \bf F1} &{\tiny \bf MCC}\\
\hline
\tiny Benchmark Industry &\tiny 63 &\tiny 39&\tiny 0 \\
\tiny Best Model in study\cite{DDR20}    &\tiny 70.1 &\tiny 61&\tiny 0.30 \\
\tiny Our  Model, VAE 3 Layers/5 Embeddings per Node &\tiny \bf78 &\tiny \bf76&\tiny \bf0.52\\
\hline
\multicolumn{4}{|c|}{\tiny  \bf Unstable Sub Period 2007-2008} \\
\hline
\tiny Benchmark Industry&\tiny 41.6 &\tiny 25&\tiny 0 \\
\tiny Best Model in study \cite{DDR20}   &\tiny 58.3   &\tiny 57&\tiny 0.29\\
\tiny Our  Model, VAE 3 Layers/5 Embeddings per Node &\tiny \bf75 &\tiny \bf75&\tiny \bf0.51\\
\hline
\multicolumn{4}{|c|}{\tiny  \bf Unstable Sub Period 2011-2012} \\
\hline
\tiny Benchmark Industry&\tiny 58.3 &\tiny 43&\tiny 0 \\
\tiny Best Model in study \cite{DDR20} &{\tiny \bf79.1}  &{\tiny \bf76}&{\tiny \bf0.61}\\
\tiny Our  Model, VAE 3 Layers/5 Embeddings per Node &\tiny 71 &\tiny 63&\tiny 0.44\\
\hline
\multicolumn{4}{|c|}{\tiny  \bf Unstable Sub Period 2015-2016} \\
\hline
\tiny Benchmark Industry&\tiny 58.3 &\tiny 43&\tiny 0\\
\tiny Best Model in study \cite{DDR20}  &\tiny 54.1   &\tiny 46& \tiny -0.06\\
\tiny Our  Model, VAE 3 Layers/5 Embeddings per Node &\tiny \bf75 &\tiny \bf75&\tiny \bf0.49\\
\hline
\end{tabular}
\end{table}
\section{Conclusion and Future Work}
In this article, we propose a framework for reconciling multiple challenges related to the use of transfer learning to represent the complex and uncertain causal process of any financial market behaviour. Then, to predict the short/medium term dynamic with an application of the monthly prediction of six of the most important US market index.\newline\\ The challenge to identify and extract knowledge from many source tasks (prediction of 235 market indexes), and then apply it to any target task that shares commonalities.\newline\\ The challenge, to transform a black-box DL framework to a hybrid white-box framework, to improve the explanatory/predictive power by creating an enabling framework for transfer learning, which allows the integration of domain knowledge at the top of the process. Thus, by using a key latent context driver or node of the causal graph of each market index as the main input. We represent each node by a relatively stable set of features which allows a reduction in the difference between the distribution of training/test sample. The framework allows merging all data, transfer feature representation from any type of financial market (Stock, Bond, Commodity, Currency, Sector, Industry and any Specific Index). Finally, the distribution obtained from latent representation of each risk factor bring the flexibility for each Portfolio Manager to deal with uncertainties and try different scenario analysis .\newline\\ 
The learning process involves two stages. The first uses an unsupervised learning process (VAE with parameters transfer) that transforms all nodes (adjacent matrix, attribute) into embedding which summarizes all the information on the underlying process of the dynamic of any market index. The second uses output of step 1 as the input for a final supervised learning algorithm on source and target tasks to predict direction of any target task.\newline\\ The experiments bring credibility to our framework. We improve three key metrics over a test period (2007-2018), and the three unstable and difficult to predict sub-periods.\newline\\
Introduce a priori knowledge in the top of the process allows: i) to try different hypotheses and preferences depending on the investment philosophy and the investment horizon; ii) to reduce the perception of black-box solution in the Asset Management Industry, and propose a hybrid solution with enough interaction between financial expert and ML expert; iii) but a potential source of criticism could be the strong dependence on an expert. The next step would involve more advanced tools and propose a more systematic process with a Knowledge Base of the dynamic of any financial market index as the main input.


\end{document}